\documentclass[a4paper,english,aps,preprint,superscriptaddress,eqsecnum]{revtex4}

\usepackage{graphicx,amsfonts,amsmath}
\usepackage{hyperref}

\DeclareMathOperator{\tr}{Tr}

\begin{document}

\title{Supersymmetric non-Abelian noncommutative Chern-Simons theory}

\author{A. F. Ferrari}
\author{ M. Gomes} 
\affiliation{Instituto de F\'\i sica, Universidade de S\~ao Paulo\\
Caixa Postal 66318, 05315-970, S\~ao Paulo, SP, Brazil}
\email{alysson,mgomes,ajsilva@fma.if.usp.br}
\author{A. Yu. Petrov}
 \affiliation{Departamento de F\'{\i}sica, Universidade Federal da Para\'{\i}ba\\
 Caixa Postal 5008, 58051-970, Jo\~ao Pessoa, Para\'{\i}ba, Brazil}
\email{petrov@fisica.ufpb.br}
\author{A. J. da Silva}
\affiliation{Instituto de F\'\i sica, Universidade de S\~ao Paulo\\
Caixa Postal 66318, 05315-970, S\~ao Paulo, SP, Brazil}
\email{alysson,mgomes,ajsilva@fma.if.usp.br}

\begin{abstract}
In this work, we study the three-dimensional non-Abelian
noncommutative supersymmetric Chern-Simons model with the $U(N)$ gauge
group. Using a superfield formulation, we prove that, for the pure
gauge theory, the Green functions are one-loop finite in any gauge, if
the gauge superpotential belongs to the fundamental representation of
$u(N)$; this result also holds when matter in the fundamental
representation is included. However, the cancellation of both ultraviolet and 
ultraviolet/infrared infrared divergences only happens in a special gauge if the coupling
of the matter is in the adjoint representation.  We also look into the
finite one-loop quantum corrections to the effective action: in the pure gauge 
sector the Maxwell together with its corresponding gauge fixing action are generated;
in the matter sector, the Chern-Simons term is generated, inducing a shift in the
classical Chern-Simons coefficient.
\end{abstract}

\maketitle

\section{Introduction}

The noncommutative Chern-Simons theory has been intensively studied
during the last years. Among the most important results obtained so
far are the vanishing of the ultraviolet/infrared (UV/IR) infrared
divergences for the pure Chern-Simons theory \cite{Schweda} and the
quantization of the Chern-Simons coefficient 
\cite{Nair,SJ,Chen,Bak}. There were also some indications that, in the
axial gauge, the pure Chern-Simons theory is actually a free
field theory \cite{Das}. Besides all these theoretical developments, a
relation of noncommutative Chern-Simons theory to the fractional
quantum Hall effect has also been pointed \cite{susskind}.

In our previous paper \cite{ourchern}, we investigated various
interactions of the noncommutative Chern-Simons field with the matter
with special concern on the existence of dangerous UV/IR
singularities. For the supersymmetric model, we found that the
cancellation of UV/IR infrared divergences occurs even when matter is
included, either in the adjoint or in the fundamental representation
(in the former case, this happens only in a particular gauge).

In this paper, we continue our line of research by looking at two
natural extensions of our previous results. First, we extend the
investigations of \cite{ourchern} to the non-Abelian case. As we shall
prove, similarly to other four- and three-dimensional noncommutative
supersymmetric gauge theories \cite{oursym1,oursym}, the cancellation
of UV/IR singularities, generated in the nonplanar part of the quantum
corrections, demands that the gauge group generators are in the
fundamental representation of the $U(N)$ group.  We recall that this is the 
same requirement found for the consistency of noncommutative
gauge models at the classical level \cite{Chaichian}. 
Second, we will calculate the finite one-loop quantum corrections to the effective
action of the supersymmetric theory (see \cite{Pani} for a similar
calculation in the non-supersymmetric case, as well as \cite{Avdeev}
for an extensive study of the commutative theory).  In particular,
this will allow us to determine the finite changes in the coefficient
of the Chern-Simons term and the structure of the effective action of
the model.

Our work is organized as follows. In the next section we demonstrate the finiteness
of the pure gauge sector if the generators are in the fundamental representation
of $u(N)$. The effect of the inclusion of matter is considered in section \ref{sec2} where we prove that finiteness in the adjoint representation also requires the gauge superfield
to be in a special gauge. The one loop calculation of the finite contributions 
to the quadratic part of the effective action is done in section \ref{sec3}. There, we prove that nonlocal Maxwell and Chern-Simons terms are generated, and the later one induces a shift in the classical Chern-Simons coefficient. In the summary, section \ref{sec4}, we present our concluding remarks.

\section{Finiteness in the pure gauge sector}
\label{sec1}

The action of the  three-dimensional noncommutative non-Abelian Chern-Simons theory with an arbitrary gauge group is $S_{total}\,=\,S_{CS}+S_{GF}+S_{FP}$~\cite{SGRS}. Here, $S_{CS}$ is the classical action

\begin{eqnarray}
S_{CS}\,=\, \frac{m}{g^2} \int d^5 z \tr \left(A^{\alpha}
*W_\alpha+\frac{i}{6}\{A^{\alpha},A^{\beta}\}\ast D_{\beta}A_{\alpha}+\frac{1}{12}
\{A^{\alpha},A^{\beta}\}*\{A_{\alpha},A_{\beta}\}\right)\,, \label{2n}
\end{eqnarray}

\noindent
$S_{GF}$ and $S_{FP}$ are the gauge fixing and Faddeev-Popov actions

\begin{eqnarray}
S_{GF}=-\frac{m}{2 g^2 \xi }\int d^5 z \tr (D^\alpha A_\alpha )\ast (D^\beta A_\beta )\
\end{eqnarray} 

\noindent
and

\begin{eqnarray}
\label{sfp}
S_{FP}=\frac{1}{2g^2}\int d^5 z \tr \left(c'\ast D^\alpha D_\alpha c+ic'*
D^\alpha [A_\alpha ,c]\right)\,,
\end{eqnarray}

\noindent
where

\begin{eqnarray}
\label{sstr}
W_\beta =\frac{1}{2}D^\alpha D_\beta A_\alpha -
\frac{i}{2}[A^\alpha ,D_\alpha A_\beta ]-
\frac{1}{6}
[A^\alpha ,\{A_\alpha ,A_\beta \}]
\end{eqnarray}

\noindent
is the superfield strength constructed from the Lie-algebra valued
spinor superpotential $A_\alpha\equiv A_\alpha^aT^a $, and the
$T^a$ are the generators of the corresponding gauge group in a given representation. 
They satisfy the normalization condition $\tr(T^aT^b)\,=\,\kappa\,\delta^{ab}$. 
Hereafter it is implicitly assumed that all commutators and
anticommutators involve both algebraic and Moyal
(anti)commutation. As usual, in this work we consider only space-space
noncommutativity, to evade unitarity problems \cite{gomis}. 

The action (\ref{2n}) is invariant under the BRST transformations 

\begin{eqnarray}
\delta A_\alpha& =&-\epsilon \nabla_\alpha c=-\epsilon(D_\alpha c +i [A_\alpha,\,
  c]), \\
\delta c& =& \epsilon c^2, \quad 
\delta  c^{\prime} \,=\, \frac{\epsilon}\xi D_\alpha A^\alpha,
\label{gt}  
\end{eqnarray}

\noindent
where $\epsilon$ is an  infinitesimal  Grassmannian gauge parameter. 
The quadratic part of the  action fixes the  gauge propagator to be

\begin{eqnarray}
\label{gaugeprop}
\label{pr1} <A^{a\alpha} (z_1)A^{b\beta}(z_2)>\,=\,
\frac{i\,g^2}{4m\kappa\,\Box}\delta^{ab}\left[
D^{\beta}D^{\alpha}+\xi D^{\alpha}D^{\beta}
\right]\delta^5(z_1-z_2)\,,
\end{eqnarray}

\noindent
where $C^{\alpha\beta}=-C_{\alpha\beta}$ is the second-rank antisymmetric Levi-Civit\`a symbol
defined with the normalization $C^{12}=i$. The ghost fields are also Lie-algebra valued and their propagator is given by

\begin{eqnarray}
\label{pr2}
<c^{\prime a}(z_1)c^b(z_2)>\,=\,-ig^2\delta^{ab}\frac{D^2}{\kappa\Box}\delta^5(z_1-z_2)\,.
\end{eqnarray}

The interaction part of the classical action in the pure gauge sector is

\begin{align}
\label{sint}
S_{int}\,= \, \frac{m}{g^2} & \int d^5 z \tr \Big(-
\frac{i}{2}A^{\alpha} \ast [A^\beta ,D_\beta A_\alpha ]-
\frac{1}{6}A^{\alpha} \ast
[A^\beta ,\{A_\beta ,A_\alpha \}] \nonumber\\
&+\frac{i}{6}\{A^{\alpha},A^{\beta}\} \ast D_{\beta}A_{\alpha}+\frac{1}{12}
\{A^{\alpha},A^{\beta}\}*\{A_{\alpha},A_{\beta}\}
\Big)\,,
\end{align}

\noindent
from which we get  the Feynman supergraphs vertices    

\begin{align}
\label{val}
&V_3=-\frac{i\,m}{3\,g^2} A^{a\alpha}(k_1)A^{b\beta}(k_2)D_{\beta}A^c_{\alpha}(k_3)
\left(A^{abc}e^{ik_2\wedge k_3}-A^{acb}e^{ik_3\wedge k_2}\right), 
\nonumber\\
&V_4=-\frac{m}{6\,g^2}  A^{a\alpha}(k_1)
A^{b\beta}(k_2)A^c_{\alpha}(k_3)A^d_{\beta}(k_4) \left[e^{i(k_1\wedge
k_2+k_3\wedge k_4)}A^{abcd}-e^{i(k_1\wedge k_3+k_2\wedge
k_4)}A^{acbd}\right], 
\end{align}

\noindent
where we have introduced the notation 

\begin{equation}
A^{ab \ldots c} \, \equiv\, \tr (T^aT^b \ldots T^c)\,.
\end{equation}

\noindent
The ghost-gauge field vertex has the following  form,

\begin{eqnarray}
V_c&=&\frac{i}{2g^2}\,D^{\alpha}c^{\prime a}(k_1)\,A^b_{\alpha}(k_2)\,c^c(k_3)\,
\left(A^{abc}e^{ik_2\wedge k_3}-A^{acb}e^{ik_3\wedge k_2}\right).
\end{eqnarray}

\begin{figure}[ht]
\begin{center} \includegraphics{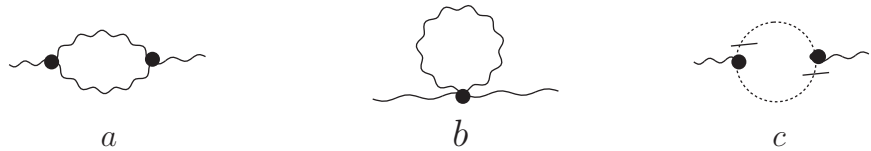}
\end{center}
\caption{Superficially linearly divergent diagrams contributing to
the two-point function of the gauge field.} \label{Fig1}
\end{figure}

We are now  in a position to study the divergences in
the one-loop quantum corrections to the effective action. 
The non-Abelian structure of the gauge group does not 
change the power counting we found for the Abelian theory \cite{ourchern}, 
so that the superficial degree of divergence for a supergraph
with $E$ external  field legs and $N_D$ covariant derivatives acting 
on those legs is $\omega=2-\frac{1}{2}E-\frac{1}{2}N_D$. 
It follows that the linear UV divergences are possible only for supergraphs with two external gauge legs. As in \cite{oursym,ours}, logarithmic divergences are absent due to symmetric integration on the loop momentum. Therefore, the complete one-loop analysis of the divergences of the (pure gauge) model demands the evaluation of the diagrams depicted in Fig.~\ref{Fig1}. 

For the sake of simplicity, we only give the details of the evaluation of the diagram \ref{Fig1}$b$. 
One starts by taking into account all contractions of two fields in the non-symmetric vertex $V_4$ in Eq.~(\ref{val}), that is to say,

\begin{align}
\Gamma_{1b}(p)=&-\frac{m}{6\, g^{2}} \,\int\left(\prod\frac{d^{3}k_{i}}{\left(2\pi\right)^{3}}\right) \, d^2\theta \,
\left\{ e^{i(k_1\wedge k_2+k_3\wedge k_4)}A^{abcd}-e^{i(k_1\wedge k_3+k_2\wedge k_4)}A^{acbd}\right\}
\times\nonumber\\
\Big[ & +A_{\alpha}^{c}\left(k_{3}\right)A_{\beta}^{d}\left(k_{4}\right)\left\langle A^{\alpha\, a}\left(k_{1}\right)A^{\beta\, b}\left(k_{2}\right)\right\rangle +A_{\alpha}^{c}\left(k_{3}\right)A_{\beta}^{b}\left(k_{2}\right)\left\langle A^{\alpha\, a}\left(k_{1}\right)A^{\beta\, d}\left(k_{4}\right)\right\rangle \nonumber\\
& +A_{\alpha}^{a}\left(k_{1}\right)A_{\beta}^{d}\left(k_{4}\right)\left\langle A^{\alpha\, c}\left(k_{3}\right)A^{\beta\, b}\left(k_{2}\right)\right\rangle +A_{\alpha}^{a}\left(k_{1}\right)A_{\beta}^{b}\left(k_{2}\right)\left\langle A^{\alpha\, c}\left(k_{3}\right)A^{\beta\, d}\left(k_{4}\right)\right\rangle \nonumber\\
& -A^{\alpha\, b}\left(k_{2}\right)A_{\alpha}^{d}\left(k_{4}\right)\left\langle A^{\beta\, a}\left(k_{1}\right)A_{\beta}^{c}\left(k_{3}\right)\right\rangle -A^{\alpha\, a}\left(k_{1}\right)A_{\alpha}^{c}\left(k_{3}\right)\left\langle A^{\beta\, b}\left(k_{2}\right)A_{\beta}^{d}\left(k_{4}\right)\right\rangle \Big]\,.\label{g1b1}
\end{align}

\noindent
The six terms in Eq.~(\ref{g1b1}) are divided in two groups of similar contributions. After some manipulations one finds, for the first four terms in the square bracket of Eq.~(\ref{g1b1}), 

\begin{align}
\Gamma_{1b}^{\left(1\right)}(p)\,=\, & -\frac{1}{24\,\kappa}\int\frac{d^{3}k}{\left(2\pi\right)^{3}} \, d^2\theta \, \left[\left(A^{abcc}-A^{bacc}\right)+2A^{abcc}-2\cos\left(2k\wedge p\right)A^{acbc}\right]\times\nonumber\\
& \left(\xi D^{\alpha}D^{\beta}+D^{\beta}D^{\alpha}\right)\delta_{11}\, A_{\alpha}^{\,a}\left(p,\theta\right)\, A_{\beta}^{\,b}\left(-p,\theta\right)\,,\label{g1b2}
\end{align}

\noindent
while, for the remaining two terms,

\begin{align}
\Gamma_{1b}^{\left(2\right)}(p)\,=\, & \frac{1}{24\,\kappa}\int\frac{d^{3}k}{\left(2\pi\right)^{3}} \, d^2\theta \, \left[-2A^{abcc}+2\cos\left(2k\wedge p\right)A^{acbc}\right]\times\nonumber\\
& \left(\xi D^{\alpha}D_{\alpha}+D_{\alpha}D^{\alpha}\right)\delta_{11}\, A^{\beta\,a}\left(p,\theta\right)\, A_{\beta}^{\, b}\left(-p,\theta\right)\,.\label{g1b3}
\end{align}

One readily realizes that the factor $\left(A^{abcc}-A^{bacc}\right)$ inside the square brackets of Eq.~(\ref{g1b2}) do not contribute for symmetry reasons. After $D$-algebra manipulations (which are trivial in this case), one can cast the contribution to the two-point vertex function of the gauge superpotential corresponding to the diagram~\ref{Fig1}$b$ as

\begin{equation}
\label{g1b}
\Gamma_{1b}(p)\,=\,\frac{1}{4\,\kappa}\left(1-\xi\right)\int\frac{d^{3}k}{\left(2\pi\right)^{3}}\,  d^2\theta \,\left[A^{abcc}-2\cos\left(2k\wedge p\right)A^{acbc}\right]
A^{\alpha\, a}\left(p,\theta\right)\, A_{\alpha}^{\, b}\left(-p,\theta\right)\,.
\end{equation}

As for the other diagrams in Fig.~\ref{Fig1}, their complete evaluation follows along the same lines but is far more intricate. Here, we just quote their divergent parts,

\begin{subequations}
\label{g1ac}
\begin{align}
\label{g1a}
\Gamma_{1a}(p)=&\frac{1}{4\kappa^{2}}\xi\int\frac{d^3k}{(2\pi)^3} \, d^2\theta \,
\left[A^{cad}A^{dbc}-A^{cad}A^{cbd}\cos(2k\wedge p)\right]
\frac{1}{k^2}A^{\alpha\,a}(p,\theta)A^b_{\alpha}(-p,\theta),\\
\label{g1c}
\Gamma_{1c}(p)=&-\frac{1}{4\kappa^{2}}\int\frac{d^3k}{(2\pi)^3} \, d^2\theta \,
\left[A^{cad}A^{dbc}-A^{cad}A^{cbd}\cos(2k\wedge p)\right]
\frac{1}{k^2}A^{\alpha\,a}(p,\theta) A^b_{\alpha}(-p,\theta)\,,
\end{align}
\end{subequations}

\noindent
postponing the detailed discussion of their finite parts to the Section~\ref{sec3}.

The planar part of $\Gamma_1 = \Gamma_{1a}+\Gamma_{1b}+\Gamma_{1c}$, being proportional to $\int\frac{d^3k}{k^2}$, exactly vanishes within the framework of
dimensional regularization, which we are implicitly assuming. As for the nonplanar part of $\Gamma_1$, which would generate a linear UV/IR infrared divergence, its cancellation is secured by the  condition involving traces of the gauge group generators, 

\begin{eqnarray}
\label{cond}
\frac{1}{\kappa}A^{cad}A^{cbd}-A^{acbc}\,=\,0\,,
\end{eqnarray}

\noindent
which is satisfied for the
$U(N)$ gauge group in the fundamental representation. Equation (\ref{cond})
is  the same condition we found for the cancellation of dangerous UV/IR infrared
singularities for the
three- and four-dimensional noncommutative super-Yang-Mills
theories \cite{oursym,oursym1}. Again, we see that the strong restriction imposed
on the choice of the gauge group at the classical level \cite{Chaichian} also plays an outstanding role to enforce the absence of singularities in the quantum theory.

As a final remark, by means of the substitution $A^{ab \ldots c} \rightarrow 1$ in Eqs.~(\ref{g1b}) and (\ref{g1ac}), we correctly reobtain the results we have found for the Abelian theory in~\cite{ourchern}.

\section{Interaction with matter}
\label{sec2}

We  will study now the consequences of the addition of matter in
the theory. Matter can interact with the Chern-Simons superfield in
two different ways. It can be  in the fundamental representation,

\begin{align}
\label{materfund}
S^{Fund}_{\rm matter}\,=\, \int d^5z &\tr \Big( \bar{\phi}(D^2-M)\phi-
\frac{i}{2}(\bar{\phi}*A^{\alpha}*D_{\alpha}\phi-D^{\alpha}\bar{\phi}*A_{\alpha}*\phi)\nonumber\\
&-\frac{1}{2}\bar{\phi}*A^{\alpha}*A_{\alpha}*\phi\Big)\,,
\end{align}

\noindent
or in the adjoint representation,

 \begin{align}
\label{matadj}
S^{Adj}_{\rm matter}\,=\, \int d^5 z & \tr \Big(\bar{\phi}(D^2-M)\phi-
\frac{i}{2}([\bar{\phi},A^{\alpha}]*D_{\alpha}\phi-D^{\alpha}\bar{\phi}\ast[A_{\alpha},\phi])\nonumber\\
&-\frac{1}{2}[\bar{\phi},A^{\alpha}]*[A_{\alpha},\phi]\Big)\,.
\end{align}

\noindent
In the presence of matter, the power counting for the supergraphs
remains the same as before, but $E$ must be considered as the total
number of external legs. Therefore, only supergraphs 
with two external legs can generate nonintegrable (linear) UV/IR
infrared divergences.

The supergraphs with a matter internal loop and external gauge legs
(see Fig. 2) are identical to the ones studied in \cite{oursym}, where
their total contribution was proved to be free of nonintegrable UV/IR
infrared divergences if and only if the condition (\ref{cond}) was
satisfied. We also note that all UV divergent terms in these diagrams cancel 
due to the relation

\begin{equation}
\label{relplanar}
\frac{1}{\kappa} A^{cad}\,A^{dac}\,=\,A^{abcc}\,,
\end{equation}

\noindent
which also happens to hold in the fundamental representation of $u(N)$. Therefore
we need to study only the graphs with external matter legs.

First, we consider matter in the fundamental representation, as specified in
Eq. (\ref{materfund}).  One can show that all the relevant diagrams
are totally planar and therefore generate no UV/IR mixing. The
cancellation of the UV divergences in these planar parts follows 
the same pattern as in the previous section.

When matter is in the adjoint representation, we find the following divergent contributions from the supergraphs depicted in Figs. \ref{Fig3}$a$ and \ref{Fig3}$b$,

\begin{align}
&\Gamma_{3a}(p)\,=\,-\frac{i\xi\,g^2}{2m\kappa^2}\int\frac{d^3k}{(2\pi)^3} \,d^2\theta\,
\frac{1}{k^2}\left[A^{cad}A^{dbc}-A^{cad}A^{cbd}\cos(2k\wedge p)\right]
\phi^a(-p,\theta)\bar{\phi}^b(p,\theta),\nonumber\\
&\Gamma_{3b}(p)\,=\,\frac{i(1-\xi)\,g^2}{2m\kappa}
\int\frac{d^3k}{(2\pi)^3}\,d^2\theta\, \frac{1}{k^{2}}
\left[A^{abcc}-A^{acbc}\cos(2k\wedge p)\right]
\phi^a(-p,\theta)\bar{\phi}^b(p,\theta).
\end{align}

\noindent
Again, the planar parts of these contributions vanish within the framework of 
dimensional regularization. The nonplanar parts would generate dangerous UV/IR mixing, but they cancel if the relation

\begin{eqnarray}
\label{cond1}
\frac{1}{\kappa}A^{cad}A^{cbd}\xi-A^{acbc}(1-\xi)\,=\,0
\end{eqnarray}

\noindent
is obeyed. The condition (\ref{cond1}) is  satisfied  only in the gauge
$\xi=1/2$ where it reproduces Eq. (\ref{cond}). Therefore, as observed before,
the cancellation only happens in the fundamental representation of $u(N)$. We remind the reader that
 the Abelian theory is also free of nonintegrable
UV/IR singularities in the gauge $\xi\,=\,1/2$ \cite{ourchern}. For a different value of $\xi$,
the matter two-point function develops an UV/IR infrared divergence,
even if the pure gauge sector is infrared safe, as we have  shown in the previous
section.

\section{Finite contribution to the two-point vertex functions}
\label{sec3}

After having demonstrated the one-loop finiteness of the model, in this section we   determine the finite one-loop quantum corrections to the quadratic part of the effective action. A  clear importance of this result is the possibility to verify whether in the quantum theory a shift in the Chern-Simons coefficient is produced.

In the pure gauge sector, the two-point vertex function of the gauge superpotential receives finite contributions only from the diagrams $a$ and $c$ in Fig.~\ref{Fig1}. The more complicated diagram is \ref{Fig1}$a$, since there are 18 different ways to contract two pairs of superfields to form the internal lines. We classify these contributions according to the distribution of supercovariant derivatives, namely 
$\Gamma_a\,=\, \Gamma_{ aI}+ \Gamma_{ aII}+ \Gamma_{ aIII}+ \Gamma_{ aIV}$ where, after simplifications of the Moyal phase factors, 

\begin{align}
\Gamma_{aI}(p)=\frac{m^{2}}{9}\, & \int\frac{d^{3}k}{\left(2\pi\right)^{3}}\, d^{2}\theta_{1}d^{2}\theta_{2}\, F_{L}^{ab}\left(k,\, p\right)\times\nonumber \\
\Big[ & +A^{\alpha\, a}\left(-p,\,\theta_{1}\right)A^{\beta^{\prime}\, b}\left(p,\theta_{2}\right)\left\langle A^{\beta}\left(1\right)\, A^{\alpha^{\prime}}\left(2\right)\right\rangle \left\langle D_{\alpha}A_{\beta}\left(1\right)\, D_{\alpha^{\prime}}A_{\beta^{\prime}}\left(2\right)\right\rangle \nonumber \\
& +A^{\alpha\, a}\left(-p,\,\theta_{1}\right)A^{\alpha^{\prime}\, b}\left(p,\theta_{2}\right)\left\langle A^{\beta}\left(1\right)\, A^{\beta^{\prime}}\left(2\right)\right\rangle \left\langle D_{\alpha}A_{\beta}\left(1\right)\, D_{\alpha^{\prime}}A_{\beta^{\prime}}\left(2\right)\right\rangle \nonumber \\
& +A^{\beta\, a}\left(-p,\,\theta_{1}\right)A^{\beta^{\prime}\, b}\left(p,\theta_{2}\right)\left\langle A^{\alpha}\left(1\right)\, A^{\alpha^{\prime}}\left(2\right)\right\rangle \left\langle D_{\alpha}A_{\beta}\left(1\right)\, D_{\alpha^{\prime}}A_{\beta^{\prime}}\left(2\right)\right\rangle \nonumber \\
& +A^{\beta\, a}\left(-p,\,\theta_{1}\right)A^{\alpha^{\prime}\, b}\left(p,\theta_{2}\right)\left\langle A^{\alpha}\left(1\right)\, A^{\beta^{\prime}}\left(2\right)\right\rangle \left\langle D_{\alpha}A_{\beta}\left(1\right)\, D_{\alpha^{\prime}}A_{\beta^{\prime}}\left(2\right)\right\rangle
\Big] \, , \label{two1}
\end{align}

\begin{align}
\Gamma_{aII}(p)=-\frac{m^{2}}{9}\, & \int\frac{d^{3}k}{\left(2\pi\right)^{3}}\, d^{2}\theta_{1}d^{2}\theta_{2}\, F_{L}^{ab}\left(k,\, p\right)\times\nonumber \\
\Big[ & +A^{\alpha\, a}\left(-p,\,\theta_{1}\right)A^{\beta^{\prime}\, b}\left(p,\theta_{2}\right)\left\langle A^{\beta}\left(1\right)\, D_{\alpha^{\prime}}A_{\beta^{\prime}}\left(2\right)\right\rangle \left\langle D_{\alpha}A_{\beta}\left(1\right)\, A^{\alpha^{\prime}}\left(2\right)\right\rangle \nonumber \\
& +A^{\alpha\, a}\left(-p,\,\theta_{1}\right)A^{\alpha^{\prime}\, b}\left(p,\theta_{2}\right)\left\langle A^{\beta}\left(1\right)\, D_{\alpha^{\prime}}A_{\beta^{\prime}}\left(2\right)\right\rangle \left\langle D_{\alpha}A_{\beta}\left(1\right)\, A^{\beta^{\prime}}\left(2\right)\right\rangle \nonumber \\
& +A^{\beta\, a}\left(-p,\,\theta_{1}\right)A^{\beta^{\prime}\, b}\left(p,\theta_{2}\right)\left\langle A^{\alpha}\left(1\right)\, D_{\alpha^{\prime}}A_{\beta^{\prime}}\left(2\right)\right\rangle \left\langle D_{\alpha}A_{\beta}\left(1\right)\, A^{\alpha^{\prime}}\left(2\right)\right\rangle \nonumber \\
& +A^{\beta\, a}\left(-p,\,\theta_{1}\right)A^{\alpha^{\prime}\, b}\left(p,\theta_{2}\right)\left\langle A^{\alpha}\left(1\right)\, D_{\alpha^{\prime}}A_{\beta^{\prime}}\left(2\right)\right\rangle \left\langle D_{\alpha}A_{\beta}\left(1\right)\, A^{\beta^{\prime}}\left(2\right)\right\rangle
\Big] \, , \label{two2}
\end{align}

\begin{align}
\Gamma_{aIII}(p)=\frac{m^{2}}{9}\, & \int\frac{d^{3}k}{\left(2\pi\right)^{3}}\, d^{2}\theta_{1}d^{2}\theta_{2}\, F_{L}^{ab}\left(k,\, p\right)\times\nonumber \\
\Big[ & +D_{\alpha}A_{\beta}^{\, a}\left(-p,\,\theta_{1}\right)D_{\alpha^{\prime}}A_{\beta^{\prime}}^{\, b}\left(p,\theta_{2}\right)\left\langle A^{\alpha}\left(1\right)\, A^{\alpha^{\prime}}\left(2\right)\right\rangle \left\langle A^{\beta}\left(1\right)\, A^{\beta^{\prime}}\left(2\right)\right\rangle \nonumber \\
& +D_{\alpha}A_{\beta}^{\, a}\left(-p,\,\theta_{1}\right)D_{\alpha^{\prime}}A_{\beta^{\prime}}^{\, b}\left(p,\theta_{2}\right)\left\langle A^{\alpha}\left(1\right)\, A^{\beta^{\prime}}\left(2\right)\right\rangle \left\langle A^{\beta}\left(1\right)\, A^{\alpha^{\prime}}\left(2\right)\right\rangle
\Big] \, , \label{two3}
\end{align}

\noindent
and

\begin{align}
\Gamma_{aIV}(p)=\frac{2m^{2}}{9}\, & \int\frac{d^{3}k}{\left(2\pi\right)^{3}}\, d^{2}\theta_{1}d^{2}\theta_{2}\, F_{L}^{ab}\left(k,\, p\right)\times\nonumber \\
\Big[ & +D_{\alpha}A_{\beta}^{\, a}\left(-p,\,\theta_{1}\right)A^{\beta^{\prime}\, b}\left(p,\theta_{2}\right)\left\langle A^{\alpha}\left(1\right)\, A^{\alpha^{\prime}}\left(2\right)\right\rangle \left\langle A^{\beta}\left(1\right)\, D_{\alpha^{\prime}}A_{\beta^{\prime}}\left(2\right)\right\rangle \nonumber \\
& +D_{\alpha}A_{\beta}^{\, a}\left(-p,\,\theta_{1}\right)A^{\alpha^{\prime}\, b}\left(p,\theta_{2}\right)\left\langle A^{\alpha}\left(1\right)\, A^{\beta^{\prime}}\left(2\right)\right\rangle \left\langle A^{\beta}\left(1\right)\, D_{\alpha^{\prime}}A_{\beta^{\prime}}\left(2\right)\right\rangle \nonumber \\
& +D_{\alpha}A_{\beta}^{\, a}\left(-p,\,\theta_{1}\right)A^{\beta^{\prime}\, b}\left(p,\theta_{2}\right)\left\langle A^{\beta}\left(1\right)\, A^{\alpha^{\prime}}\left(2\right)\right\rangle \left\langle A^{\alpha}\left(1\right)\, D_{\alpha^{\prime}}A_{\beta^{\prime}}\left(2\right)\right\rangle \nonumber \\
& +D_{\alpha}A_{\beta}^{\, a}\left(-p,\,\theta_{1}\right)A^{\alpha^{\prime}\, b}\left(p,\theta_{2}\right)\left\langle A^{\beta}\left(1\right)\, A^{\beta^{\prime}}\left(2\right)\right\rangle \left\langle A^{\alpha}\left(1\right)\, D_{\alpha^{\prime}}A_{\beta^{\prime}}\left(2\right)\right\rangle
\Big] \, . \label{two4}
\end{align}

\noindent
The Moyal phase factor common to all these diagrams is given by

\begin{equation}
\label{flab}
F_{L}^{ab}\left(k,\, p\right)\,=\,\left(A^{cad}A^{dbc}-A^{cad}A^{cbd}\cos2k\wedge p\right) \,,
\end{equation}

\noindent
and, in a slight abuse of notation, factors like $< A_{\alpha}(1)A_{\beta}(2) >$ in Eqs.~(\ref{two1})-(\ref{two4}) mean the gauge superpropagator in Eq.~(\ref{gaugeprop}) without the $\delta^{ab}$, which has already been used to simplify the phase factor $F_{L}^{ab}\left(k,\, p\right)$.

Finally, the graph in Fig. \ref{Fig1}$c$ yields the ghost contribution,

\begin{align}
\Gamma_{c}(p)=\frac{1}{2}\,\int\frac{d^{3}k}{\left(2\pi\right)^{3}} & \, d^{2}\theta_{1}d^{2}\theta_{2}\, F_{L}^{ab}\left(k,\, p\right)A_{\beta}^{\, a}\left(-p,\,\theta_{1}\right) A_{\beta}^{\, b}\left(p,\theta_{2}\right)\times\nonumber \\
& \left[D_{1}^{\alpha}D^{2}\delta^{2}\left(\theta_{1}-\theta_{2}\right)\right]
\left[D^{2}D_{2}^{\beta}\delta^{2}\left(\theta_{1}-\theta_{2}\right)\right]
\, . \label{mlett:c1}
\end{align}

Only the contributions $\Gamma_{aI}$ and $\Gamma_{c}$ have divergent parts, which were already discussed in Section~\ref{sec2}. For the remaining of this section, we restrict ourselves to the discussion of the finite parts of $\Gamma_{a}$ and $\Gamma_{c}$. To evaluate them, one has to perform rather lengthy $D$-algebra manipulations, thus generating lots of terms. These were calculated independently by hand and also by means of a computer program designed for quantum superfield calculations~\cite{susymath}. 

We do not quote all the details of the calculation, but we will describe the necessary steps using $\Gamma_{aI}$ as a typical example of the contributions we have evaluated. First of all, using $D$-algebra manipulations, we can cast $\Gamma_{aI}$ as follows,

\begin{align}
\Gamma_{aI}^{Fin}(p)&\,=\,-\frac{1}{144} \int\frac{d^3k}{(2\pi)^3} \,d^2 \theta\,
\frac{F_{L}^{ab}\left(k,\, p\right)}{k^2(p-k)^2}
(k_{\alpha\delta}C_{\beta\gamma}+k_{\beta\gamma}C_{\alpha\delta})\times\nonumber\\
\Big[ & (D^{\beta}D^{\gamma}+\xi D^{\gamma}D^{\beta})
A^{\alpha\,a}(p,\theta)A^{\delta\,b}(-p,\theta)+
(D^{\alpha}D^{\delta}+\xi D^{\delta}D^{\alpha})
A^{\beta\,a}(p,\theta)A^{\gamma\,b}(-p,\theta)+\nonumber\\
+ &(D^{\beta}D^{\delta}+\xi D^{\delta}D^{\beta})
A^{\alpha\,a}(p,\theta)A^{\gamma\,b}(-p,\theta)+
(D^{\alpha}D^{\gamma}+\xi D^{\gamma}D^{\alpha})
A^{\beta\,a}(p,\theta)A^{\delta\,b}(-p,\theta) \Big],
\end{align}

\noindent
where the superscript $Fin$ is to stress that we are quoting only the
finite part. Next, we introduce the definition

\begin{equation}
\label{defi}
I^{ab}\left(p\right)\,=\,\int\frac{d^{3}k}{\left(2\pi\right)^{3}}\frac{F_{L}^{ab}\left(k,\, p\right)}{k^{2}\left(k+p\right)^{2}}\,,
\end{equation}

\noindent
and verify that

\begin{eqnarray}
\label{thei}
\int\frac{d^3k}{(2\pi)^3}\frac{F_{L}^{ab}\left(k,\, p\right)}{k^2(p-k)^2}k_{\alpha\beta}
=\frac{p_{\alpha\beta}}{2} I^{ab}(p)\,,
\end{eqnarray}

\noindent
to write

\begin{align}
\Gamma_{aI}^{Fin}=&\frac{1}{36}\int \frac{d^3p}{(2\pi)^3} \, d^2\theta\,  I^{ab}(p) \,
\Big[p^2 A^{\alpha\,a} (p,\theta)A_{\alpha}^b (-p,\theta)(1+\xi)\nonumber\\
&+
p_{\alpha\beta}D^2A^{\alpha\,a}(p,\theta)A^{\beta\,b}(-p,\theta)(1-\xi)
\Big]\,.
\end{align}

\noindent
The final form of $\Gamma_{aI}^{Fin}$ and is obtained by means of the identities~\cite{SGRS}

\begin{align} 
\int \frac{d^3p}{(2\pi)^3}\,d^2\theta\,
p^2A^{\gamma\,a}(p,\theta)A_{\gamma}^b(-p,\theta) &= 
\int \frac{d^3p}{(2\pi)^3}\,d^2\theta\, \left[-2\left(L_{Maxw}\right)^{ab}+2\left(L_{MGF}\right)^{ab}\right],\nonumber\\ 
\int \frac{d^3p}{(2\pi)^3} \,d^2\theta\, p_{\beta\gamma}
D^2A^{\beta\,a}(p,\theta)A^{\gamma\,b}(-p,\theta) &= 
\int \frac{d^3p}{(2\pi)^3} \,d^2\theta\, \left[2\left(L_{Maxw}\right)^{ab}+2\left(L_{MGF}\right)^{ab}\right]\,,
\end{align}

\noindent
where

\begin{equation}
\label{finali}
\left(L_{Maxw}\right)^{ab}=\frac{1}{2}W^{\alpha\,a}_0 W_{0\alpha }^b,\,\quad\, 
\left(L_{MGF}\right)^{ab}=\frac{1}{4}D^{\beta}A_{\beta}^a(p,\theta)D^2D^{\gamma}A_{\gamma}^b(-p,\theta)
\end{equation}

\noindent
are the Maxwell (``linearized Yang-Mills'') and the gauge-fixing Lagrangians of the three-dimensional SYM theory, respectively. Also, $W^{\beta\,a}_0 \,\equiv\, \frac{1}{2}D^{\alpha}D^{\beta}A_{\alpha}^a$ 
is the linear part of the superfield strength defined in~(\ref{sstr}).

Using the steps just described, we obtained the finite parts of  $\Gamma_{a}$ and $\Gamma_{c}$ as follows,

\begin{eqnarray}
\Gamma_{aI}^{Fin}&=&\frac{1}{9} \int \frac{d^3p}{(2\pi)^3} \, d^2\theta \, I^{ab}(p) \,
\Big\{\xi \left(L_{Maxw}\right)^{ab} - \left(L_{MGF}\right)^{ab}\Big\}\,,\nonumber\\
\Gamma_{aII}^{Fin}&=&\frac{1}{36} \int \frac{d^3p}{(2\pi)^3} \, d^2\theta \, I^{ab}(p) \,
\Big\{\xi(\xi+4)\left[\left(L_{Maxw}\right)^{ab}-\left(L_{MGF}\right)^{ab}\right]\Big\}\,,\nonumber\\
\Gamma_{aIII}^{Fin}&=&-\frac{1}{36} \int \frac{d^3p}{(2\pi)^3} \, d^2\theta \, I^{ab}(p) \,
\Big\{(1-\xi)\left[4 \xi \left(L_{Maxw}\right)^{ab} + (1-\xi) \left(L_{MGF}\right)^{ab}\right]\Big\}\,, \nonumber\\
\Gamma_{aIV}^{Fin}&=&-\frac{1}{9} \int \frac{d^3p}{(2\pi)^3} \, d^2\theta \, I^{ab}(p) \,
\Big\{ (1-\xi)\left[ \xi \left(L_{Maxw}\right)^{ab} +(1+\frac{1}{2}\xi)\left(L_{MGF}\right)^{ab}\right]\Big\}\,, \nonumber\\
\Gamma_{c}^{Fin}&=&\frac{1}{4} \int \frac{d^3p}{(2\pi)^3} \, d^2\theta \, I^{ab}(p) \,
\Big\{\left(L_{Maxw}\right)^{ab}-\left(L_{MGF}\right)^{ab}\Big\}\,.
\end{eqnarray}

\noindent
As a result, the total pure gauge contribution to the two-point gauge superpotential function is 

\begin{eqnarray}
\Gamma_{2\,\textrm{gauge}}&=&\frac{1}{4} \int \frac{d^3p}{(2\pi)^3} \, d^2\theta \, I^{ab}(p) \,
\Big[ (1 + \xi^2) \left(L_{Maxw}\right)^{ab}-2 \left(L_{MGF}\right)^{ab}\Big].
\end{eqnarray}

Therefore, we find that both nonlocal Maxwell and the gauge-fixing actions are generated by quantum corrections, and there is no gauge in which any of them vanish. Note that the Chern-Simons term is not generated here.

\begin{figure}[ht]
\begin{center}\includegraphics{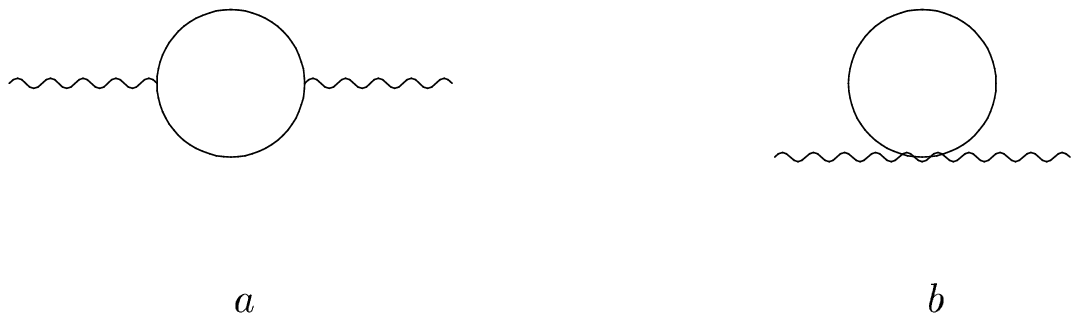}
\end{center}
\caption{Superficially linearly divergent diagrams contributing to
the two-point function of the gauge field: matter sector} \label{Fig2}
\end{figure}

\begin{figure}[ht]
\begin{center}\includegraphics{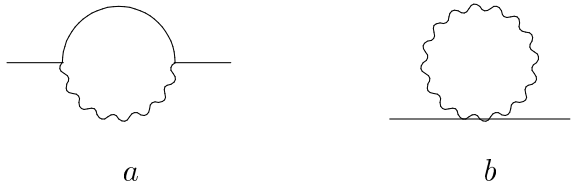}
\end{center}
\caption{Diagrams contributing to the two-point function of the matter field.} \label{Fig3}
\end{figure}

The only source for the Chern-Simons term as a quantum correction is the coupling to the matter (see Fig.~\ref{Fig2}) in which case we obtained the following contribution to the two-point gauge superpotential function~\cite{cpn},

\begin{align}
\label{stot}
\Gamma_{2\,\textrm{matter}}\,=\,&-\frac{1}{2} \int  \frac{d^3p}{(2\pi)^3}  \frac{d^3k}{(2\pi)^3} \,d^2\theta\, J^{ab}(k,p) \,
(k_{\gamma\beta}+MC_{\gamma\beta}) \nonumber\\
&\Big[(D^2A^{\gamma\,a}(-p,\theta)) A^{\beta\,b}(p,\theta)
+\frac{1}{2} D^{\gamma}D^{\alpha}A_{\alpha}^a(-p,\theta) A^{\beta\,b}(p,\theta)
\Big]\,,
\end{align}

\noindent
where

\begin{equation}
\label{final}
J^{ab}(k,p) \, = \,  \frac{1}{(k^2+M^2)[(k-p)^2+M^2]}
\left[ \begin{array}{c} A^{cad}A^{dbc} \\ 2 F_L^{ab}(k,\,p) \end{array} \right] \,.
\end{equation}

\noindent
In Eq.~(\ref{final}), the upper (lower) row corresponds to the fundamental (adjoint) representation for the matter field. An equivalent form for $\Gamma_{2matter}$ is

\begin{equation}
\label{t2p1}
\Gamma_{2\,\textrm{matter}} \,=\, \frac{1}{4} \int \frac{d^3 k}{(2\pi)^3} \, d^2\theta f^{ab}(p)
\left[ W^{\alpha\,a}_0 W^b_{0\alpha }+ 2M W_{0}^{\alpha\,a} A^b_\alpha \right],
\end{equation}

\noindent
where

\begin{equation}
f^{ab}(p) \, = \, \int \frac{d^3 k}{(2\pi)^3} J^{ab}(k,p)\,.
\end{equation}

\noindent
We stress that the simplicity of Eq.~(\ref{t2p1}) depends on the relation~(\ref{relplanar}), which enforces the cancellation of additional finite terms that would appear in Eq.~(\ref{t2p1}) if the generators $T^a$ were not in the fundamental representation of $u(N)$.

The above results show explicitly the generation of nonlocal Maxwell and Chern-Simons terms. From Eq.~(\ref{t2p1}), we find  that the radiative corrections to the original Chern-Simons coefficient $m/{g^2}$ turn out to be

\begin{equation}
\frac{1}{2} M f^{ab}(0) \, = \, \frac{N}{16\pi}\varepsilon(M)\, 
\left[  \begin{array}{c}  A^{cad}A^{dbc} \\  2 F_L^{ab}(p=0)  \end{array} \right]\,.
\end{equation}

\noindent
Here, $\varepsilon$ is the sign function. Recalling the completeness relation satisfied by the generators of the $U(N)$ group in the fundamental representation, 
$\left(T^a\right)_{ij}\left(T^a\right)_{kl}\,=\,\kappa \delta_{il} \delta_{jk}$, and that the value of $\kappa$ in this representation is $1/2$, one obtains

\begin{equation}
\label{csquant}
\frac{1}{2} M f^{ab}(0) \, = \, \frac{N}{128\pi}\varepsilon(M)\, 
\left[  \begin{array}{c}  \delta^{ab} \\  2 \left( \delta^{ab}-\delta^{a0}\delta^{b0} \right) \end{array} \right]\,.
\end{equation}

\noindent
For the matter in the fundamental representation, the above result is similar to the ones found in the literature for nonsupersymmetric theories~\cite{SJ,Chen}, where a shift proportional to $N\,\varepsilon(M)$ was also found. On the other hand, when matter is in the adjoint representation, the shift only appears for the non-Abelian components of the gauge superpotential. This is consistent with the absence of this shift in the Abelian theory, as can be seen from our calculations in~\cite{ourchern}.

\section{Summary}
\label{sec4}
We studied the quantum dynamics of the noncommutative supersymmetric
Chern-Simons theory with an arbitrary gauge group, in the one-loop
approximation. Our first result consists in the fact that the
cancellation of the nonintegrable (linear) UV/IR infrared divergences
occurs only if the generators of the gauge group are in the
fundamental representation of $U(N)$. This condition is
the same as the one found for the four-dimensional super-Yang-Mills
\cite{oursym1} and the three-dimensional super-Yang-Mills \cite{oursym}
theories in spite of the radical difference between their superfield
formulations.

We have also calculated the finite one-loop corrections to the gauge superpotencial two-point function, and we verified that a nonlocal Maxwell and the corresponding gauge fixing terms are generated from the pure gauge sector. We note that none of them vanish in any gauge. From the matter radiative corrections, besides the nonlocal Maxwell, a Chern-Simons term is also induced. If the matter is in the fundamental representation, there is a shift in the classical Chern-Simons coefficient proportional to $N\,\varepsilon(M)$. Interestingly enough, for the matter in the adjoint representation, the Chern-Simons coefficient receives quantum corrections only for the non-Abelian components of the gauge superpotential. 


\section{Acknowledgments} 
This work was partially supported by Funda\c{c}\~{a}o de Amparo 
\`{a} Pesquisa do Estado de S\~{a}o Paulo (FAPESP) and Conselho 
Nacional de Desenvolvimento Cient\'{\i}fico e Tecnol\'{o}gico (CNPq).
The work of A. F. F. was supported by FAPESP,  project 04/13314-4. 
A. Yu. P. has been supported by CNPq-FAPESQ, DCR program (CNPq project 350400/2005-9).


\end{document}